%% file: main.tex
\documentclass{ceurart}

\usepackage{booktabs}
\usepackage{tikz}
\usetikzlibrary{positioning,fit,backgrounds}
\usepackage{graphicx}
\usepackage{xcolor}

\definecolor{ontcrm}{HTML}{1B1B1B}
\definecolor{ontinf}{HTML}{7B2D8B}
\definecolor{ontdig}{HTML}{0E7C7B}
\definecolor{ontsci}{HTML}{3A3FA0}
\definecolor{famTemporal}{HTML}{3783A8}
\definecolor{famActors}{HTML}{72005F}
\definecolor{famConceptual}{HTML}{C78F15}
\definecolor{famPhysical}{HTML}{801401}
\definecolor{famAppellations}{HTML}{68810C}
\definecolor{famPlaces}{HTML}{206D33}
\definecolor{famTypes}{HTML}{F88C34}
\definecolor{famNone}{HTML}{D8D8D8}

\begin{document}

\copyrightyear{2026}
\copyrightclause{Copyright for this paper by its authors.
  Use permitted under Creative Commons License Attribution 4.0
  International (CC BY 4.0).}

\conference{Sci-K 2026: 6th International Workshop on Scientific Knowledge:
  Representation, Discovery, and Assessment, co-located with ISWC 2026,
  October 25--26, 2026, Bari, Italy}

\title{The CIDOC CRM Periodic Table: Documentation You Can Test to Support the
  Adoption of Scientific Research Ontologies for the Humanities}

%% Single-anonymous review: the author is named, not anonymised.
%% ORCID, email and affiliation taken from the CITATION.cff of release 1.0.0.
\author[1,2]{Remo Grillo}[%
orcid=0000-0003-3585-0198,
email=remo.grillo@unibo.it,
url={https://remogrillo.github.io/cidoc-crm_periodic_table/},
]
\address[1]{Alma Mater Studiorum --- Universit\`a di Bologna, Bologna, Italy}
\address[2]{I Tatti --- The Harvard University Center for Italian Renaissance
  Studies, Florence, Italy}

\begin{abstract}
  The CIDOC CRM family is not only a schema for museum collections. Its extensions
  formalise the apparatus through which research knowledge is produced and
  justified: CRMsci models scientific observation and measurement, CRMdig the
  provenance of digital processes, CRMinf inference and belief. Together they
  constitute a schema for research data, and the obstacle to their adoption is
  documentary rather than theoretical: four normative documents, four code
  conventions, and eighty-nine typed links crossing document boundaries. This paper
  presents the CIDOC CRM Periodic Table, a single-page browser rendering 131 classes
  and 480 property entries from CIDOC CRM 7.1.3, CRMsci 3.2, CRMdig 5.0 and
  CRMinf 1.2.1 as one navigable surface, with two inference facilities: a one-step
  triple builder admitting only the class--property combinations licensed by declared
  domains, ranges and the subclass closure, and a multi-step path finder enumerating
  property chains between two classes. Because all loaded models are
  merged into one code-keyed dictionary before the hierarchy closures are computed,
  both resolve chains that cross model boundaries; we document the addition of
  CRMsci and the three cross-model tests it made possible, and report preliminary
  uptake signals from 56 months of analytics, with the reading they support.
\end{abstract}


\maketitle

\section{Introduction}
\label{sec:intro}

A standard is adopted through its documentation, and a standard that grows by
extension fragments its documentation along the same seams: what a modeller needs to
know ends up distributed across the very modules that make the standard extensible.
The CIDOC Conceptual Reference Model~\cite{crm713}, standardised as ISO 21127, is
accompanied by a family of extensions that, taken together, describe how
empirical claims come to be: CRMsci~\cite{crmsci32} defines itself as ``a global
schema for integrating metadata about scientific observation, measurements and
processed data in descriptive and empirical sciences such as life sciences,
geology, geography, archaeology, cultural heritage conservation and others in
research IT environments and research data libraries''; CRMdig~\cite{crmdig50}
models the provenance of digital processes; CRMinf~\cite{crminf121} models
inference, argumentation and belief. A researcher who wants to say what was
observed, by which instrument, through which digital transformation, and on what
grounds the resulting statement is believed, already has a normative vocabulary
for all four moves.

What that researcher does not have is a workable way to read it. Nine years ago,
in a peer-reviewed venue, Guillem and Bruseker argued that CIDOC CRM adoption in
concrete practice is limited because ontologies are ``considered too complicated
and abstract for application in real life scenarios''~\cite{guillem2017game}.
Since then the specification has grown rather than shrunk. Covering the four
models discussed here means reading four separate documents that declare 131
classes and 480 property entries, use four disjoint code conventions
(\texttt{E}/\texttt{P}, \texttt{S}/\texttt{O}, \texttt{D}/\texttt{L},
\texttt{I}/\texttt{J}), and connect to each other through 89 \texttt{subClassOf}
and \texttt{subPropertyOf} declarations that point out of the document one is
currently reading. The barrier is documentary, and it grows with every extension the
family gains.

This paper describes the CIDOC CRM Periodic Table~\cite{periodictable}, a
client-side tool%
\footnote{Live interface: \url{https://remogrillo.github.io/cidoc-crm_periodic_table/};
  source code: \url{https://github.com/RemoGrillo/cidoc-crm_periodic_table}; archived
  release: \url{https://doi.org/10.5281/zenodo.21821748} (concept DOI). The interface is
  a static page: serving it locally needs no build step and no dependencies, and the
  repository README gives the commands for that and for regenerating the data files from
  the official RDFS sources.} that addresses this barrier, and reports on the recent addition of
CRMsci. Its contributions are:

\begin{itemize}
  \item a single-page interface presenting CIDOC CRM 7.1.3 with CRMsci 3.2,
        CRMdig 5.0 and CRMinf 1.2.1 as one surface, with per-model filtering and
        shareable deep links (Section~\ref{sec:design});
  \item two inference facilities --- a constrained one-step triple builder and a
        bounded multi-step path finder --- whose coverage and non-coverage we state
        exactly, and a demonstration that the family's cross-model links are
        resolved structurally rather than case by case, through three tests enabled
        by the CRMsci import (Section~\ref{sec:inference});
  \item preliminary uptake signals over 56 months, with the caveats a reader needs
        in order not to over-read them (Section~\ref{sec:uptake}).
\end{itemize}

\section{Background: the CRM family as research-data infrastructure}
\label{sec:background}

Read individually, each extension looks like a specialist concern. Read together,
they cover the lifecycle of an empirical claim. Take a mundane research workflow:
a sample is taken, it is measured with an instrument, the instrument writes a file,
and someone draws a conclusion. Every step is already declared, across three of the
four models: \texttt{S2 Sample Taking -O5 removed-> S13 Sample};
\texttt{S21 Measurement -O24 measured-> S13}, admissible because \texttt{S13}
descends from \texttt{S15 Observable Entity}, the declared range, with
\texttt{S21 -O39 observed dimension-> E54 Dimension} for the value;
\texttt{D11 Digital Measurement Event} $\sqsubseteq$ \texttt{S21}, CRMdig
subclassing CRMsci, with \texttt{D11 -L20 has created-> D9 Data Object}; and
\texttt{S5 Inference Making} $\sqsubseteq$ \texttt{I1 Argumentation}, CRMsci
subclassing CRMinf, with \texttt{I1 -J2 concluded that-> I2 Belief}. This is a
schema for research data, not for catalogue records, which is what Beretta's SDHSS
ecosystem argues from the side of the social sciences and
humanities~\cite{beretta2024sdhss}. Its provenance half overlaps with what
PROV-O~\cite{provo} models generically; CRMdig differs in committing to the same
event-centric ontology as the observation and inference halves --- which is
precisely why the four documents have to be read together.

The decisive point for tooling is that these models are \emph{interlocked, not
parallel}: they are not four self-contained documents, each readable on its own, but
documents that declare entities whose superclass or superproperty is defined in another
one. \texttt{S5 Inference Making} is declared a subclass of CRMinf's
\texttt{I1 Argumentation}; \texttt{S27 Observation} likewise; CRMdig's
\texttt{D11 Digital Measurement Event} is a subclass of CRMsci's
\texttt{S21 Measurement}, and \texttt{L1 digitized} a subproperty of CRMsci's
\texttt{O24 measured}. Counting from the machine-readable releases, the four
models declare 89 links whose target lives in a different document
(Appendix~\ref{app:coverage}): 31 from CRMdig, 30 from CRMsci and 28 from CRMinf.
Ten of CRMsci's point at CRMinf rather than at base CRM, and three of CRMdig's
point at CRMsci.

The practical consequence is that reading one specification is never enough, and
the workflow above shows it concretely. A modeller who opens CRMdig alone sees that
\texttt{D11} creates a \texttt{D9}, and misses that \texttt{D11} \emph{is} a
measurement and therefore already carries \texttt{O24 measured} and
\texttt{O39 observed dimension}; a modeller who opens CRMsci alone cannot see that
the properties of \texttt{I1 Argumentation} apply to \texttt{S5}. The official HTML
declarations, published per model and per version, faithfully mirror the documents
--- and therefore mirror the fragmentation too. This is the gap the interface
targets, and it is a documentation problem of the kind the community meets whenever
scholarly knowledge is represented in a modular standard: comparable
infrastructures for scholarly knowledge, such as the Open Research Knowledge
Graph~\cite{jaradeh2019orkg}, invest heavily in making their model legible to
contributors for the same reason.

\section{Related work}
\label{sec:related}

Three genres of resource already address parts of this problem, and it is worth
being precise about which part (Table~\ref{tab:compare}).

\emph{Specification renderings} are faithful to their input and inherit its
boundaries: the CRM-SIG's machine-generated declarations give one HTML page per model
per version, and generated documentation such as WIDOCO~\cite{garijo2017widoco} or
visualisers such as WebVOWL~\cite{lohmann2015webvowl} do the same for arbitrary
vocabularies. In all of them a reference into another model stays a string.

\emph{Ontology management.} OntoME is the closest neighbour --- a collaborative
environment handling multiple namespaces, profiles and versioning, which lets users
align their own classes to CIDOC CRM and compatible models~\cite{beretta2021ontome}
and does resolve the family as one graph. Its centre of gravity is authoring and
curation, one page per entity; ours is reading a whole model at once and testing
what it licenses.

\emph{General-purpose tooling.} One can load all four RDFS files into an editor such
as Prot\'eg\'e~\cite{musen2015protege} with an OWL reasoner behind it, and in terms of
raw capability that suffices. The difference is one of purpose. An editor supports
\emph{building and verifying} an ontology: its unit of work is the axiom, its
question whether a model is consistent. The task here is downstream and far more
common --- \emph{applying} a standardised ontology one did not design, translating
``this sample was measured, and here is the file the instrument wrote'' into its
codes. An editor supplies the vocabulary but no help in choosing within it, and
asking ``which properties may hold between a sample and a data object'' costs a
reasoning environment and a formalised query; here it costs a URL. Documentation and
modelling support are usually separate artefacts; for a modular standard they should
not be.

Three neighbours sit outside the table. Mountantonakis et al.\ catalogue which
\emph{datasets} use which classes and properties~\cite{mountantonakis2023portals}:
instance-level evidence, where we offer schema-level navigation. Linked Art narrows
the model instead, publishing a profile over a subset of CIDOC CRM
7.1.3~\cite{linkedart} --- a subset helps the modeller who accepts its scope,
navigation helps the one who cannot. Closest in spirit at this workshop, Tietz et
al.\ treat cultural-heritage material as research data and build discovery services
over it~\cite{tietz2025cto}: the same premise, applied to the data rather than the
schema. Tools we could not verify are deliberately not cited.

\begin{table}[t]
  \centering
  \caption{What each genre offers for the four capabilities this paper claims.
    Compiled from each resource's own public documentation, consulted 4 August
    2026: ``--'' means the capability is not among the documented features, not
    that it is unattainable. \emph{One surface} = every loaded model visible in a
    single view; \emph{cross-document closure} = a subclass chain whose target is
    declared in another model's document is resolved; \emph{constrained triple} =
    admissible classes and properties are derived from declared domains, ranges and
    the subclass closure; \emph{chain search} = property paths between two given
    classes are enumerated. The ontology editor is listed as an upper bound on raw
    capability rather than as a documentation resource: what separates it from the last
    row is purpose, not expressive power.}
  \label{tab:compare}
  \begin{tabular}{lcccc}
    \toprule
    & One & Cross-document & Constrained & Chain \\
    & surface & closure & triple & search \\
    \midrule
    Official HTML declarations         & per model & --          & --        & -- \\
    Generated docs (WIDOCO)            & per model & --          & --        & -- \\
    Graph visualisers (WebVOWL)        & per model & --          & --        & -- \\
    Ontology management (OntoME)       & --        & \checkmark  & --        & -- \\
    Ontology editor + reasoner         & --        & \checkmark  & \checkmark & partial \\
    \textbf{Periodic Table}            & \checkmark & \checkmark & \checkmark & \checkmark \\
    \bottomrule
  \end{tabular}
\end{table}

\section{Design}
\label{sec:design}

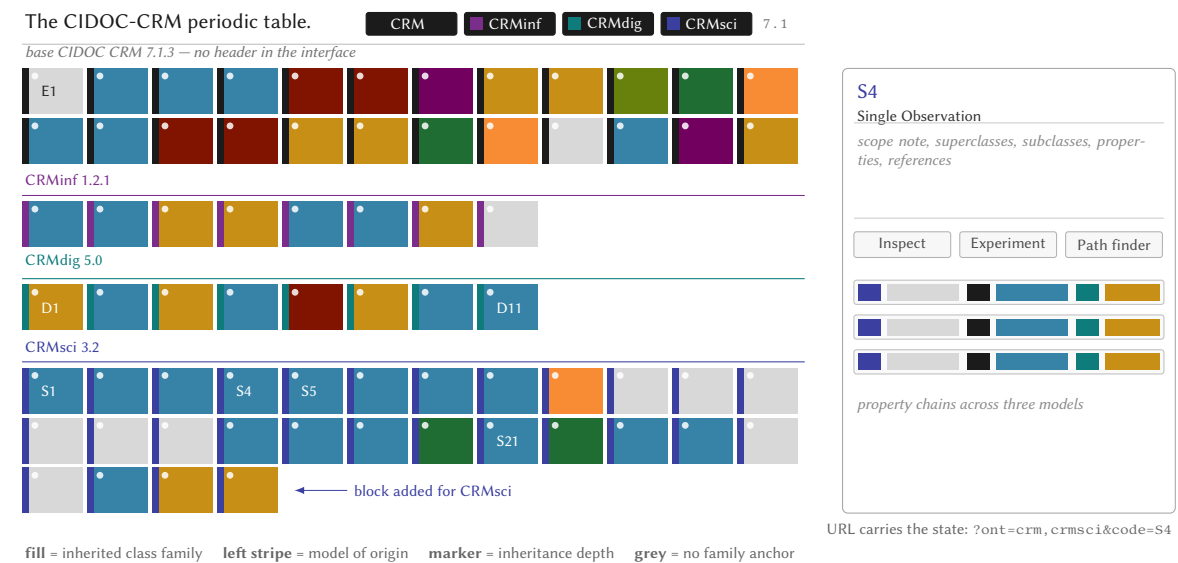
\begin{figure}[!ht]
  \centering
  \input{fig-layout}
  \caption{Layout logic. Classes of every enabled model share one scrolling
    surface: base CIDOC CRM first, then one contiguous block per extension,
    each introduced by a full-width header in the model's provenance hue
    (violet CRMinf, teal CRMdig, indigo CRMsci). Cell fill encodes the
    inherited class family, the left stripe encodes the model, the small marker
    encodes inheritance depth, and grey marks a class that descends from no
    family anchor. Toggling a model hides its cells without unloading it, so
    hierarchy closures stay intact. The inspector on the right flips to reveal
    the two inference panels of Section~\ref{sec:inference}.}
  \label{fig:layout}
\end{figure}

\begin{figure}[h]
  \centering
  \fbox{\includegraphics[width=0.9\linewidth]{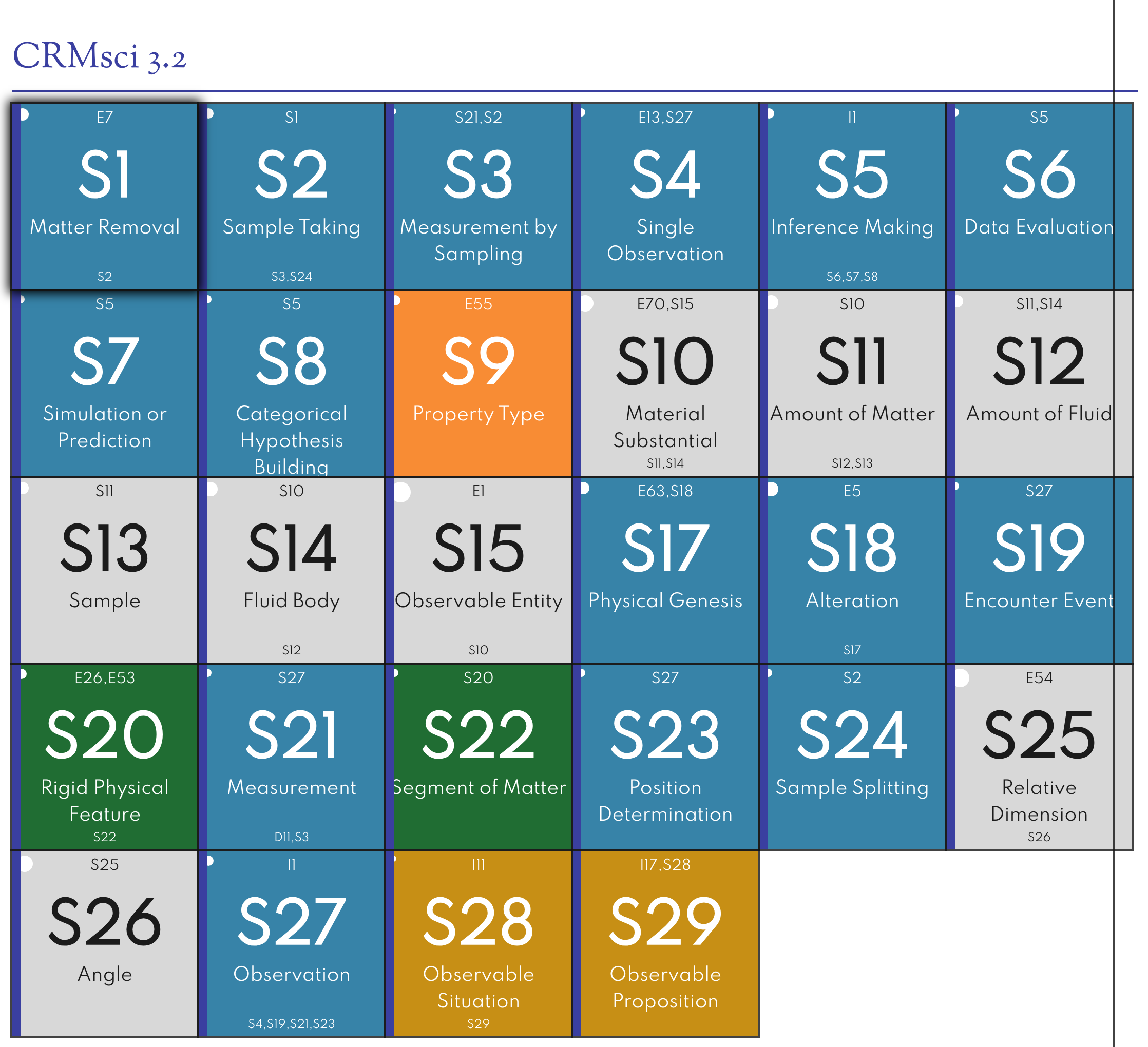}}
  \caption{The CRMsci block as it appears in the table after the import: 28
    classes under a header in the model's hue, each cell carrying an indigo
    provenance stripe, the codes of its immediate superclasses, a marker for
    inheritance depth, and the fill of the class family it inherits. Blue cells
    descend from \texttt{E4 Temporal Entity}, green from \texttt{E53 Place}, ochre
    from \texttt{E28 Conceptual Object}, orange from \texttt{E55 Type}; grey marks
    the classes that descend from no anchor. Note the cross-model superclass codes:
    \texttt{I1} above \texttt{S5} and \texttt{S27}, \texttt{I11} above
    \texttt{S28}. Cells are shown at full opacity; in use, the interface dims
    those unrelated to the current selection.}
  \label{fig:crmsciblock}
\end{figure}

The interface renders every class of every enabled model as a cell on a single
scrolling page, and every property as a line in a companion view
(Figure~\ref{fig:layout} schematises that surface and
Figure~\ref{fig:crmsciblock} shows one block of it as rendered). Four decisions do
most of the work.

\textbf{Placement follows the code.} A cell's code prefix determines
which model it belongs to and whether it is a class or a property; the four code
spaces are disjoint, so a code denotes exactly one entry across the loaded family.
Cells then flow inline and wrap, so the grid reflows to the width available instead
of being pinned to fixed columns. Each model occupies one contiguous block, introduced by a
full-width header in the model's own hue, so provenance is legible without a
legend.

\textbf{Colour is inherited through the hierarchy.} Seven anchor classes ---
\texttt{E4 Temporal Entity}, \texttt{E18 Physical Thing}, \texttt{E39 Actor},
\texttt{E28 Conceptual Object}, \texttt{E41 Appellation}, \texttt{E53 Place},
\texttt{E55 Type} --- define seven colour families, and every class takes the colour
of the anchor it descends from through the \texttt{subClassOf} closure. Nothing is
hard-coded per class, so adding a model costs no colour work: 20 of CRMsci's 28
classes were coloured as intended the moment they were loaded. Grouping CRM's classes
into colour-coded families is not our idea: we inherit it, along with the association
between family and hue, from the CIDOC CRM Game~\cite{guillem2017game,crmgame2021}.
The anchors, the values, the derivation of colour by inheritance, and the second
colour axis marking which model an entry comes from are ours; no graphic material
from the game was reused.

\textbf{Inheritance depth is legible on the cell.} Each cell carries the codes of its
immediate superclasses and a marker that shrinks with the length of its longest
inheritance path, separating \texttt{E1} from a class eight steps below it.

\textbf{The URL carries the state.} The query string of the browser URL ---
\texttt{?ont=crm,crmsci}, \texttt{?code=S4}, \texttt{?path=S13,D9} --- is read on load
and rewritten on interaction, so any class, filter combination or path query is a
citable link, and a deep link silently enables the model that owns its entry. These are
view parameters and not identifiers: an entity's URI remains the one its ontology
declares, carried on the cell and on the copy action, though it deserves to be shown in
the detail panel too.

Accommodating CRMsci exercised exactly these mechanisms. It uses \texttt{S} for
classes and \texttt{O} for properties, a convention shared with neither base CRM nor
the other extensions, so it needed a registry entry, two code-prefix declarations, a
data file and a provenance hue --- and nothing else. The hue was chosen by measurement
rather than taste: indigo (\texttt{\#3A3FA0}) sits 37 degrees of hue from its nearest
neighbour among the two model hues and the seven family colours, at 8.8:1 contrast on
white.

\section{Inference support}
\label{sec:inference}

A documentation browser is easy to build; letting a modeller assemble a triple and
see what the specification licenses is the distinctive part, and the place where
over-claiming is most tempting, so we state the boundaries plainly.

\textbf{One step.} The user picks a domain class, a property and a range class in
any order, and the table dims every entry that the choice has excluded. The rule is
subsumption over declared domains and ranges: a class is admissible in a slot if it
is the declared domain or range of the property, or a subclass of it through the
transitive \texttt{subClassOf} closure; a property is admissible if the classes
already chosen fall within its declared domain and range in the same sense. The
completed triple is read back as an English sentence and as parsable Turtle. The
Turtle is built at instance level, with blank nodes and only the \texttt{@prefix}
declarations the triple actually uses, because CRM properties hold between instances
and a triple asserted between classes would be meaningless.

\begin{figure}[t]
  \centering
  \fbox{\includegraphics[width=0.80\linewidth]{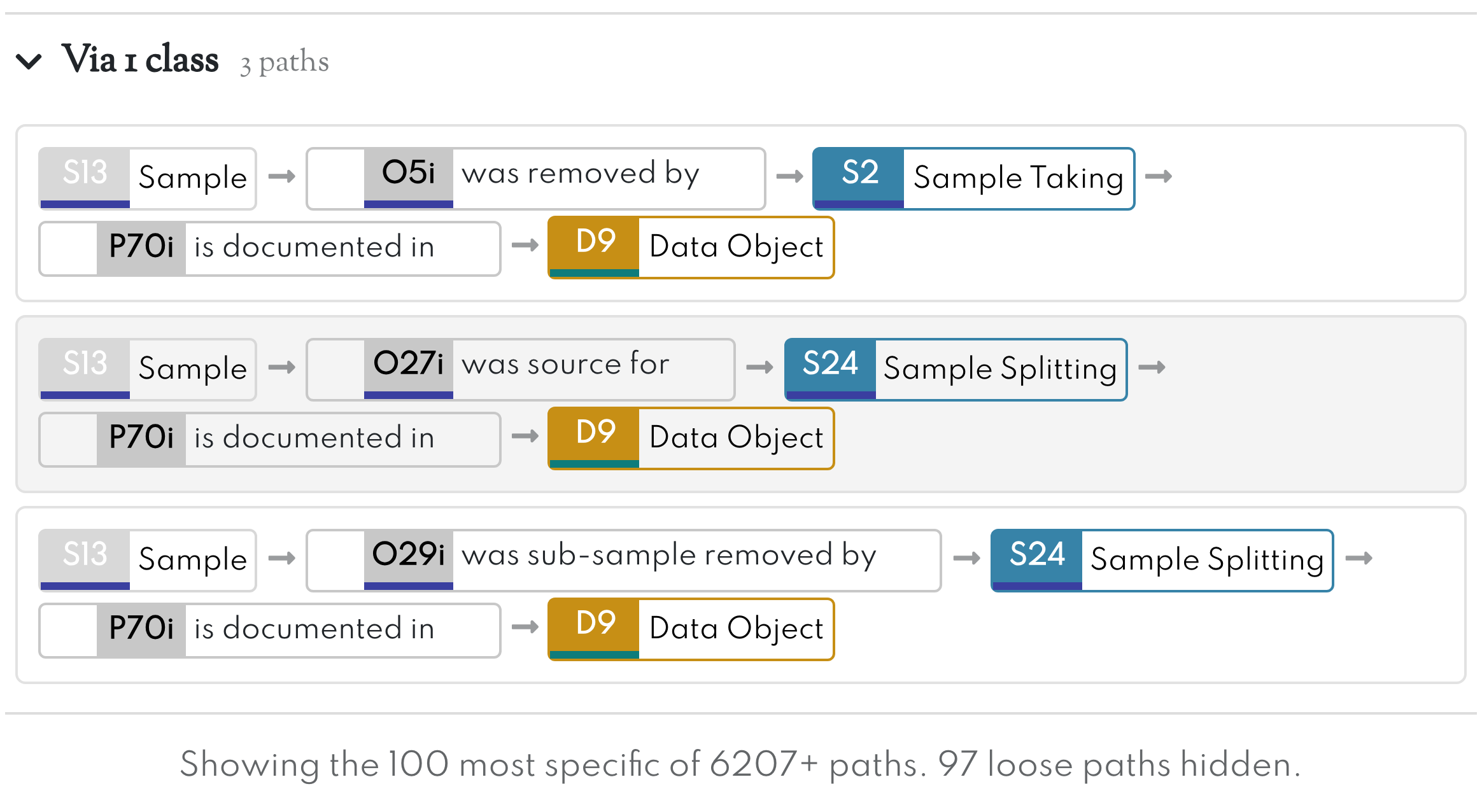}}
  \caption{Multi-step output for \texttt{S13 Sample} to \texttt{D9 Data Object} at
    three steps, sound chains only. All three cross three models: CRMsci classes
    (indigo stripe) reach a CRMdig class (teal stripe) through the base-CRM property
    \texttt{P70i}, resolved by inherited domains and not by any
    namespace-specific rule. The footer is reproduced from the interface,
    truncation notice included.}
  \label{fig:pathfinder}
\end{figure}

\textbf{Multiple steps.} The path finder takes two classes and enumerates property
chains between them, using a bounded bidirectional breadth-first search over the
domain/range graph, with subsumption applied at every hop: a property is available
on a class if the class or any of its superclasses is the declared domain. Results
are grouped by chain length and ranked by the specificity of the properties
involved. The bounds are what make this tractable: each side expands at most half the
requested depth, every level is beam-limited to its 600 most specific states and the
answer set to 100, so the work per query grows with those caps and the branching of the
domain/range graph rather than exponentially with depth, and any truncation is reported
on screen. The search runs synchronously in the browser over the whole loaded family;
it is the caps, not the size of the family, that keep it interactive. An optional \emph{loose} mode admits hops that are valid only if the
intermediate instance belongs to a more specific class than the chain guarantees ---
the downcasts that are common in real modelling --- and such chains are dashed and
labelled in the output, never silently mixed with sound ones
(Figure~\ref{fig:pathfinder}).

\textbf{Cross-model chains are structural.} Every loaded model is merged into one
dictionary keyed by bare code, and the subclass and subproperty closures are
computed after the merge, over the union. Nothing in the search knows about
namespaces; a chain crosses a model boundary the same way it crosses any other
edge. The CRMsci import let us test this on the three cases that exist in the
data; each is a deep link, so each is re-enterable
(\texttt{?ont=crm,crminf,crmdig,crmsci\&path=}). First, CRMsci to base CRM:
\texttt{S4,E53} yields 11 sound chains within two steps, five of them touching
three models. Second, CRMsci to CRMinf: \texttt{S5,I2} yields exactly one sound
chain, \texttt{S5 -J2-> I2}, and it exists only because \texttt{S5 Inference
Making} inherits the domain of a CRMinf property through
\texttt{S5 $\sqsubseteq$ I1 Argumentation}. This is the case a reader of CRMsci
alone cannot see, and it is the example we would put in front of a sceptic. Third,
CRMdig to CRMsci, \texttt{D11,S13}: before the import, \texttt{D11}'s superclass
\texttt{S21} was an inert label and no chain could pass through it; after it,
\texttt{D11 -O24-> S13} is found and flagged loose, while \texttt{S13,D9} returns
\texttt{S13 -O5i-> S2 -P70i-> D9}, which spans CRMsci, base CRM and CRMdig
soundly.

\textbf{What is verified, and what is not.} Two things were checked and are
reproducible from the script distributed with this paper: the loaded entity counts
against the official class and property declarations (28 classes and 39 forward
properties for CRMsci 3.2, plus 38 materialised inverses), and the transitive
subclass closure across model boundaries. The three path results reported above
were obtained from the deployed interface and are stated here with the queries that
produce them, so a reader can re-enter each one --- they are documented, not
scripted. We did \emph{not} verify the
implementation systematically against the full set of entailments the
specifications declare, and we do not claim to. Concretely, the following are
absent by design and should not be inferred from the word ``inference'':
\texttt{subPropertyOf} entailment is not applied, so asserting a subproperty does
not surface its superproperty --- a real limitation here, given that CRMsci
declares 40 subproperty links, 10 of them into base CRM; inverse properties are
present because the RDFS materialises them, not because an \texttt{owl:inverseOf}
rule is applied; property transitivity, cardinality and the CRM quantifier
statements are not modelled; shortcut properties are not held equivalent to their
full paths; and there is no consistency checking, so the tool prevents an
inadmissible triple rather than diagnosing one. The multi-step search is
incomplete by construction --- depth is capped at four steps, each frontier is
beam-limited, and at most 100 results are returned --- and the interface reports
truncation on screen. A ``no path found'' answer is therefore not a proof of
absence, and we would rather say so than let a reader assume otherwise.

In the vocabulary of research assessment, the two facilities make a representation's
\emph{coverage} countable (Appendix~\ref{app:coverage}) and its
\emph{interpretability} testable: what the schema licenses becomes something one can
question instead of reconstruct. That is not a quality metric, and nothing here
substitutes for one.

\section{Preliminary uptake signals}
\label{sec:uptake}

The tool carries a Google Analytics 4 property whose record begins on 1 December
2021; the repository is older. Over the 56 months to 4 August 2026 the property
recorded 2\,552 active users and 94\,078 events, with a mean engagement time of 540
seconds per active user, and 60\,517 page views on its main title. Users were
attributed to 654 distinct cities, 111 of them with five or more users, and the
author's own institution accounts for 82 sessions out of 9\,222: use is
overwhelmingly external.

Two patterns in the referrers are more informative than the totals
(Appendix~\ref{app:referrers} lists the hosts). Teaching circulation is visible:
302 sessions arrive from course environments, including Google Classroom and the
Moodle instances of five universities. Professional circulation is visible too: 224
sessions arrive from links opened inside working documents and meetings --- Word
documents served through \texttt{officeapps.live.com}, Teams chats, Webex and Zoom.
Community referral comes chiefly from GitHub (302 sessions) and
\texttt{linked.art} (97 sessions). Separately, three page titles that appear
in the property but in no version of this repository account for 2\,587 views by
150 users; the titles announce further extensions, so the most economical reading is
an independently deployed derivative that inherited the analytics tag. We cannot
verify that from the outside, we exclude those views from every other figure in this
section, and we mention them only because a reused tool is worth more than a
consulted one.

These data support statements about reach and about the kinds of setting in which
the tool is opened. They support no statement whatsoever about learning, usability
or didactic effectiveness: there has been no user study, and none is claimed. Four
caveats bound even the reach claims. First, only GA4's default bot exclusion is in
force, so sophisticated headless crawlers may be counted. Second, ad blockers
systematically depress these counts, and this audience is unusually likely to run
them. Third, and most easily misread: because the interface routes by code,
inspecting a class generates a page view, so the view count is a proxy for depth of
exploration and not for distinct visits --- a reviewer who opens the site will
confirm this in seconds. Fourth, the property was never filtered for the author's
own development traffic; 31 sessions from \texttt{localhost} and self-referrals are
still in the data.

\section{Limitations and roadmap}
\label{sec:limitations}

The tool has never been evaluated with users, and the analytics above are not a
substitute. Coverage of the family is partial: CRMgeo, CRMarchaeo and CRMba are
not loaded, and references into them remain inert labels. The inference
facilities cover the fragment described in Section~\ref{sec:inference} and no
more. There is no automated test suite. Two interface decisions belong here as well.
The layout assumes a desktop viewport: the premise of the tool is a whole model in view
at once, which is desk work, and it does not lay out for small screens. And the colour
families, derived and documented in Section~\ref{sec:design}, have no legend in the
interface itself --- the per-model chips double as one for provenance, the seven family
hues have no equivalent.

\textbf{How far this generalises.} Within the CRM family the answer is settled by
construction: a further model costs a registry entry, its code prefixes, a data file and
a hue, as CRMsci did; models with two-letter prefixes --- CRMgeo's \texttt{SP},
CRMarchaeo's \texttt{AP} --- additionally need the code pattern widened, a contained
change. Beyond the family the boundary is not in the code but in an assumption about the
domain. The table rests on a convention that is universal in the CRM family and general
nowhere: that every class and property carries a short mnemonic code of one letter and a
number. That code decides the cell, the owning model and the class/property split --- and
it is declared nowhere in the RDF, there being no property that states it, so it is
recovered heuristically from the local name of the URI. A vocabulary without such codes
could still have the cross-document closure, the constrained triple builder and the path
finder, which need only RDFS; it could not have the periodic table. We would therefore
claim generalisation to modular, code-bearing standards, and no further.

The roadmap follows from that list, in order of what would change most: a teaching
study in a graduate course, where the tool is already being used informally;
systematic verification of the implemented entailments against the normative
declarations, beginning with \texttt{subPropertyOf}; the remaining
compatible extensions, for which the CRMsci import is now the template; and two small
debts to the reader, a legend for the colour families and the entity URI shown in the
detail panel.

\section{Conclusion}
%% The body proper ends with this section; the label measures it (see NOTES.md).

The CIDOC CRM family already offers a normative vocabulary for observation, digital
provenance and inference --- for the apparatus by which research data is produced and
justified. What it does not offer is a way to read four interlocking documents at
once and check what they license together, and that gap is not a detail of
presentation: it is where the cost of adopting the standard is actually paid.

The Periodic Table takes the modest position that this is a tooling problem rather
than a modelling one. Its design result is small enough to state in a sentence ---
merge every loaded model into one dictionary keyed by bare code, then compute the
hierarchy closures over the union --- and its consequence is that cross-model
inspection stops being a special case and becomes the default. Adding CRMsci
exercised precisely that: a registry entry, two code prefixes, a data file and a
colour, after which chains cut at the document boundary resolved on their own, and
the same four steps --- with the code pattern widened where a model's prefixes run
to two letters --- are what it would take to bring in CRMgeo, CRMarchaeo or CRMba.

Two limits travel with the claim. The inference facilities implement a fragment ---
subclass transitivity and domain/range applicability --- and we say which entailments
are absent, because a tool that answers ``what may I say about this?'' is trusted, and
trust must not be borrowed against. And it has never been evaluated with users: 56
months of traffic establish that people reach it, not that it helps them. What we would draw from this is more general than CIDOC CRM: modular standards
distribute their meaning across documents and leave the reader to recompose it; where
that recomposition is mechanical, the documentation can perform it.
\label{lastbodypage}

\section*{Declaration on Generative AI}

During the preparation of this work, the author used Claude (Anthropic) in order to:
Grammar and spelling check. Further, the author used Claude (Anthropic) during the
development of the CIDOC CRM Periodic Table in order to: Generate code, confined to bug
fixing in an existing codebase. After using these tools, the author reviewed and edited
the content as needed and takes full responsibility for the content of this publication.

No generative AI system was used to generate or analyse research data, or to draft the
scientific content of this manuscript. The interface described here was written by hand
and its development began before such assistants were available; AI assistance came late
and was limited to the bug fixing stated above. All bibliographic references were
verified against primary sources, and every quantitative statement was recomputed from
the source data.

\bibliography{references}

\appendix

\section{Full coverage table}
\label{app:coverage}

Table~\ref{tab:fullcoverage} reports the coverage figures quoted in the body
split by property direction, together with the cross-model link counts and the
alternative CIDOC CRM 6.2.1 base, which the tool also offers and for which no
extension has a compatible machine-readable release. All numbers are produced by
\texttt{count\_coverage.py} from the JSON files the interface loads at runtime;
those files are generated from the official RDFS releases by the converter in the
repository, so the chain from specification to figure is reproducible.

\begin{table}[h]
  \centering
  \caption{Coverage and cross-model links, counted from the machine-readable
    releases. Forward properties exclude the inverse entries materialised in the
    RDFS. CRMsci 3.2 declares 39 forward properties and 38 inverses because
    \texttt{O30 determined position} has a literal range and therefore no
    inverse.}
  \label{tab:fullcoverage}
  \begin{tabular}{llrrrl}
    \toprule
    Model & Ver. & Cls. & Fwd. & Inv. & Links out of the document \\
    \midrule
    CIDOC CRM & 7.1.3 & 75 & 164 & 144 & --- \\
    CRMsci    & 3.2   & 28 &  39 &  38 & 10 + 10 to CRM; 4 + 6 to CRMinf \\
    CRMdig    & 5.0   & 14 &  22 &  21 & 8 + 20 to CRM; 1 + 2 to CRMsci \\
    CRMinf    & 1.2.1 & 14 &  27 &  25 & 4 + 24 to CRM \\
    \midrule
    \textbf{Total} & & \textbf{131} & \textbf{252} & \textbf{228} & \textbf{89} \\
    \midrule
    CIDOC CRM & 6.2.1 & 84 & 153 & 133 & alternative base, no extensions \\
    \bottomrule
  \end{tabular}
\end{table}

Link counts are given as \emph{subclass + subproperty}. The 89 cross-document
links are what makes the merged view more than a convenience: each one is a fact
about the family that no single specification document states in full.

CIDOC CRM 6.2.1 declares more classes than 7.1.3 (84 against 75) because the 7.x
revisions deprecated and merged several classes of the 6.x series; the figure is
not a counting error, and the tool offers 6.2.1 only as an alternative base, with
no extension attached.

\section{Referring hosts}
\label{app:referrers}

Sessions by referring host, from the same export, grouped as in
Section~\ref{sec:uptake}. Course environments: \texttt{classroom.google.com} 158,
\texttt{moodle.lmu.de} 98, \texttt{e.fhs.upr.si} 15, \texttt{online.uni-marburg.de}
and \texttt{ddk-data.online.uni-marburg.de} 15, \texttt{virtuale.unibo.it} 8,
\texttt{moodle.univie.ac.at} 6, \texttt{confluence.uni-muenster.de} 1,
\texttt{redmine.ihpan.edu.pl} 1. Documents, chat and meetings:
\texttt{statics.teams.cdn.office.net} 118, the four regional
\texttt{*-word-edit.officeapps.live.com} hosts 50 in total,
\texttt{mail.google.com} 15, \texttt{unito.webex.com} 13,
\texttt{docs.google.com} 9, \texttt{zimbra.univ-lille.fr} 8,
\texttt{zimbra.inha.fr} 5, \texttt{web.skype.com} 3, \texttt{st1.zoom.us} 2,
\texttt{teams.microsoft.com} 1. Community: \texttt{github.com} 302,
\texttt{linked.art} 97, \texttt{t.co} 8, \texttt{intarch.ac.uk} 5,
\texttt{researchgate.net} 1. Search engines 3\,464, of which
\texttt{google} 3\,352. AI assistants: \texttt{chatgpt.com} 22,
\texttt{copilot.microsoft.com} 3. Author's institution:
\texttt{dh.itatti.harvard.edu} 82. Development traffic, unfiltered:
\texttt{remogrillo.github.io} 22, \texttt{localhost:8000} 4,
\texttt{127.0.0.1:5500} 3, \texttt{localhost:5500} 1. Direct, with no referrer,
4\,676. Total 9\,222 sessions from 44 distinct sources.

\clearpage
\section{Additional figures}
\label{app:figures}

\begin{figure}[h]
  \centering
  \fbox{\includegraphics[width=0.8\linewidth]{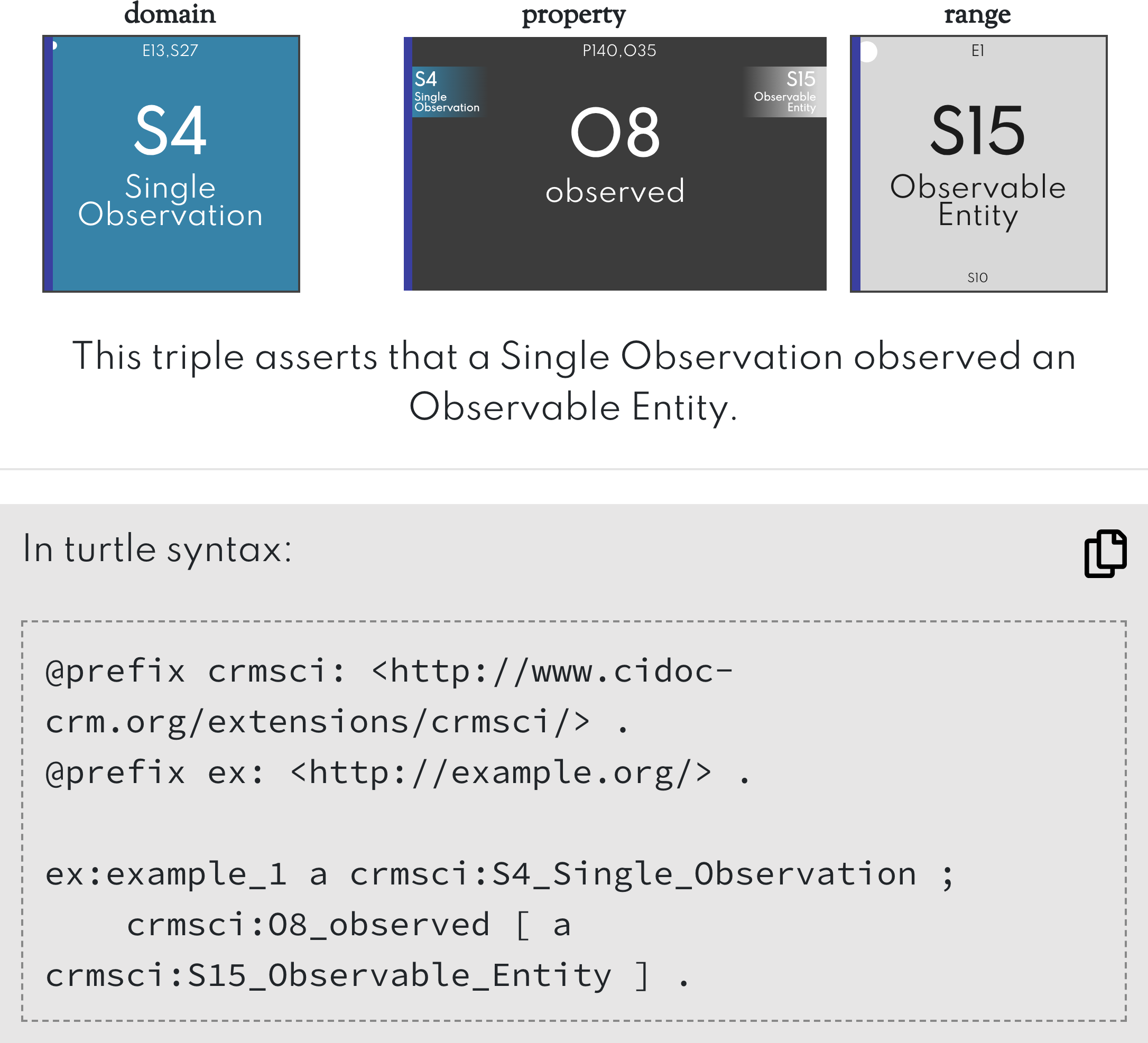}}
  \caption{One-step output for \texttt{S4 Single Observation},
    \texttt{O8 observed}, \texttt{S15 Observable Entity}: the natural-language
    read-back and the instance-level Turtle with the prefixes the triple
    actually uses. Only 6 of the 131 loaded classes are admissible in the range
    slot for \texttt{O8} --- \texttt{S15} and its five subclasses --- and the
    table dims the rest while the choice is open.}
  \label{fig:experiment}
\end{figure}

\end{document}

%% file: fig-layout.tex
% Figure 1: layout logic of the CIDOC CRM Periodic Table (schematic, vector).
% Colours are the tool's own values, declared in main.tex.
% Vertical rhythm: cell height 0.60, row pitch 0.66, block separation 0.44.
\begingroup
\def\cw{0.80}   % cell width
\def\ch{0.60}   % cell height
\def\gp{0.06}   % gap between cells

% #1 x, #2 y (top-left), #3 fill, #4 model stripe.
% Coordinates are evaluated first, so callers may pass arithmetic.
\newcommand{\ptcell}[4]{%
  \pgfmathsetmacro{\cx}{#1}%
  \pgfmathsetmacro{\cy}{#2}%
  \fill[#3] (\cx,\cy) rectangle ++(\cw,-\ch);
  \fill[#4] (\cx,\cy) rectangle ++(0.09,-\ch);
  \pgfmathsetmacro{\dx}{\cx+0.17}%
  \pgfmathsetmacro{\dy}{\cy-0.10}%
  \fill[white,opacity=0.85] (\dx,\dy) circle (0.045);
}
% labelled cell: #5 = code, #6 = text colour
\newcommand{\ptcelll}[6]{%
  \ptcell{#1}{#2}{#3}{#4}%
  \pgfmathsetmacro{\lx}{#1+0.21}%
  \pgfmathsetmacro{\ly}{#2-\ch/2}%
  \node[anchor=west,font=\tiny\sffamily,text=#6] at (\lx,\ly) {#5};
}
% one row of unlabelled cells: #1 y, #2 list of index/colour pairs
\newcommand{\ptrow}[3]{%
  \foreach \i/\f in {#2} {\ptcell{\i*(\cw+\gp)}{#1}{\f}{#3}}%
}
% block header: #1 y of the rule, #2 label, #3 colour, #4 right end
\newcommand{\blockhead}[4]{%
  \pgfmathsetmacro{\hy}{#1+0.09}%
  \node[anchor=south west,font=\tiny\sffamily,text=#3] at (0,\hy) {#2};
  \draw[#3,line width=0.4pt] (0,#1) -- (#4,#1);
}

\begin{tikzpicture}[every node/.style={inner sep=1pt}]

  \def\W{10.35}     % right edge of the classes surface
  \def\CL{10.85}    % inspector card, left
  \def\CR{15.25}    % inspector card, right

  % ---------- toolbar ----------
  \node[anchor=north west,font=\scriptsize\sffamily,text=ontcrm]
    at (0,0.04) {The CIDOC-CRM periodic table.};
  \foreach \i/\c/\n in {0/ontcrm/CRM, 1/ontinf/CRMinf, 2/ontdig/CRMdig, 3/ontsci/CRMsci} {
    \pgfmathsetmacro{\bx}{4.55+\i*1.30}
    \draw[ontcrm,fill=ontcrm,rounded corners=1pt] (\bx,0) rectangle ++(1.20,-0.30);
    \fill[\c] (\bx+0.08,-0.07) rectangle ++(0.16,-0.16);
    \node[anchor=west,font=\tiny\sffamily,text=white] at (\bx+0.28,-0.15) {\n};
  }
  \node[anchor=west,font=\tiny\ttfamily,text=black!55] at (9.75,-0.15) {7.1};
  \draw[black!25] (0,-0.42) -- (\W,-0.42);

  % ---------- base CRM: no header in the interface ----------
  \node[anchor=south west,font=\tiny\itshape,text=black!55] at (0,-0.68)
    {base CIDOC CRM 7.1.3 --- no header in the interface};
  \ptcelll{0}{-0.74}{famNone}{ontcrm}{E1}{black!85}
  \ptrow{-0.74}{1/famTemporal, 2/famTemporal, 3/famTemporal, 4/famPhysical,
    5/famPhysical, 6/famActors, 7/famConceptual, 8/famConceptual,
    9/famAppellations, 10/famPlaces, 11/famTypes}{ontcrm}
  \ptrow{-1.40}{0/famTemporal, 1/famTemporal, 2/famPhysical, 3/famPhysical,
    4/famConceptual, 5/famConceptual, 6/famPlaces, 7/famTypes, 8/famNone,
    9/famTemporal, 10/famActors, 11/famConceptual}{ontcrm}

  % ---------- extension blocks ----------
  \blockhead{-2.42}{CRMinf 1.2.1}{ontinf}{\W}
  \ptrow{-2.50}{0/famTemporal, 1/famTemporal, 2/famConceptual, 3/famConceptual,
    4/famTemporal, 5/famTemporal, 6/famConceptual, 7/famNone}{ontinf}

  \blockhead{-3.52}{CRMdig 5.0}{ontdig}{\W}
  \ptcelll{0}{-3.60}{famConceptual}{ontdig}{D1}{white}
  \ptrow{-3.60}{1/famTemporal, 2/famConceptual, 3/famTemporal, 4/famPhysical,
    5/famConceptual, 6/famTemporal}{ontdig}
  \ptcelll{7*(\cw+\gp)}{-3.60}{famTemporal}{ontdig}{D11}{white}

  \blockhead{-4.62}{CRMsci 3.2}{ontsci}{\W}
  \ptcelll{0}{-4.70}{famTemporal}{ontsci}{S1}{white}
  \ptrow{-4.70}{1/famTemporal, 2/famTemporal}{ontsci}
  \ptcelll{3*(\cw+\gp)}{-4.70}{famTemporal}{ontsci}{S4}{white}
  \ptcelll{4*(\cw+\gp)}{-4.70}{famTemporal}{ontsci}{S5}{white}
  \ptrow{-4.70}{5/famTemporal, 6/famTemporal, 7/famTemporal, 8/famTypes,
    9/famNone, 10/famNone, 11/famNone}{ontsci}
  \ptrow{-5.36}{0/famNone, 1/famNone, 2/famNone, 3/famTemporal, 4/famTemporal,
    5/famTemporal, 6/famPlaces}{ontsci}
  \ptcelll{7*(\cw+\gp)}{-5.36}{famTemporal}{ontsci}{S21}{white}
  \ptrow{-5.36}{8/famPlaces, 9/famTemporal, 10/famTemporal, 11/famNone}{ontsci}
  \ptrow{-6.02}{0/famNone, 1/famTemporal, 2/famConceptual, 3/famConceptual}{ontsci}

  % ---------- inspector / inference card ----------
  \draw[black!35,fill=white,rounded corners=2pt] (\CL,-0.74) rectangle (\CR,-6.62);
  \node[anchor=north west,font=\scriptsize\sffamily,text=ontsci] at (11.00,-0.90) {S4};
  \node[anchor=north west,font=\tiny\sffamily,text=black!80] at (11.00,-1.24) {Single Observation};
  \draw[black!20] (11.00,-1.46) -- (15.10,-1.46);
  \node[anchor=north west,font=\tiny\itshape,text=black!45,align=left,text width=3.9cm]
    at (11.00,-1.58) {scope note, superclasses, subclasses, properties, references};
  \draw[black!20] (11.00,-2.72) -- (15.10,-2.72);
  \foreach \i/\t in {0/{Inspect}, 1/{Experiment}, 2/{Path finder}} {
    \pgfmathsetmacro{\bx}{11.00+\i*1.40}
    \draw[black!30,fill=black!3,rounded corners=1pt] (\bx,-2.90) rectangle ++(1.28,-0.34);
    \node[font=\tiny\sffamily,text=black!70] at (\bx+0.64,-3.07) {\t};
  }
  \foreach \i in {0,1,2} {
    \pgfmathsetmacro{\y}{-3.54-\i*0.46}
    \draw[black!25,fill=black!2,rounded corners=1pt] (11.00,\y) rectangle (15.10,\y-0.32);
    \fill[ontsci] (11.06,\y-0.05) rectangle ++(0.30,-0.22);
    \fill[famNone] (11.44,\y-0.05) rectangle ++(0.95,-0.22);
    \fill[ontcrm] (12.50,\y-0.05) rectangle ++(0.30,-0.22);
    \fill[famTemporal] (12.88,\y-0.05) rectangle ++(0.95,-0.22);
    \fill[ontdig] (13.94,\y-0.05) rectangle ++(0.30,-0.22);
    \fill[famConceptual] (14.32,\y-0.05) rectangle ++(0.72,-0.22);
  }
  \node[anchor=north west,font=\tiny\itshape,text=black!45] at (11.00,-5.06)
    {property chains across three models};

  % ---------- callouts ----------
  \draw[ontsci,-latex,line width=0.3pt] (4.30,-6.32) -- (3.60,-6.32)
    node[anchor=west,font=\tiny,text=ontsci] at (4.35,-6.32) {block added for CRMsci};
  \node[anchor=north east,font=\tiny,text=black!60] at (\CR,-6.70)
    {URL carries the state: \texttt{?ont=crm,crmsci\&code=S4}};
  \node[anchor=north west,font=\tiny,text=black!60,align=left] at (0,-7.02)
    {\textbf{fill} = inherited class family \quad
     \textbf{left stripe} = model of origin \quad
     \textbf{marker} = inheritance depth \quad
     \textbf{grey} = no family anchor};

\end{tikzpicture}
\endgroup